\documentclass[conf]{new-aiaa}
\usepackage[utf8]{inputenc}

\usepackage{graphicx}
\usepackage{amsmath}
\usepackage[version=4]{mhchem}
\usepackage{siunitx}
\usepackage{longtable,tabularx}
\newcommand{\vMin}{{v}_\mathit{min}} 
\newcommand{\vMax}{{v}_\mathit{max}} 
\newcommand{\aComfort}{{a}_\mathit{comf}} 
\newcommand{\aEmergency}{{a}_\mathit{max}} 
\newcommand{\bComfort}{{b}_\mathit{comf}} 
\newcommand{\bEmergency}{{b}_\mathit{min}} 
\newcommand{\xf}{{x}_f} 
\newcommand{\vf}{{v}_f} 
 
\newcommand{\xr}{{x}_r} 
\newcommand{\vr}{{v}_r} 
 
\newcommand{\xii}{{x}_i} 
\newcommand{\vi}{{v}_i} 
\newcommand{\ai}{{a}_i} 
\newcommand{\corridorLength}{L} 
\newcommand{\tfEnter}{{t}_\mathit{f}}
\newcommand{\trEnter}{{t}_\mathit{r}}
\newcommand{\tiEnter}{{t}_\mathit{i}}
\newcommand{\durationF}{{\tau}_\mathit{f}}
\newcommand{\durationR}{{\tau}_\mathit{r}}
\newcommand{\durationI}{{\tau}_\mathit{i}}
\newcommand{\delay}{\delta} 
\newcommand{\dRSS}{\mathit{dRSS}} 
\newcommand{\dNMAC}{\mathit{dNMAC}} 

\newtheorem{prob}{Problem}
\newtheorem{thm}{Theorem}
\newtheorem{cor}{Corollary}

\usepackage{subfigure}
\usepackage{esint}
\usepackage{xcolor}

\title{Safe Arrival Scheduling at Constraint Waypoints\\in UAM Corridors}

\author{Sasinee Pruekprasert\footnote{Researcher, NEC – AIST AI Cooperative Research Laboratory, Artificial Intelligence Research Center, s.pruekprasert@aist.go.jp.}}
\affil{National Institute of Advanced Industrial Science and Technology, Koto, Tokyo, 135-0064, Japan}

\author{Shinji Nakadai
\footnote{CEO, Intent Exchange, Inc., nakadai@intent-exchange.com.}}
\affil{Intent Exchange, Inc., Bunkyo, Tokyo, 113-0023, Japan}

\begin{document}

\thispagestyle{empty}

\noindent This is a preprint of a paper published in the
Proceedings of the AIAA SCITECH 2025 Forum.

\noindent Copyright \copyright\ 2025 by Sasinee Pruekprasert and Shinji Nakadai.

\noindent The published version is available at:
\texttt{https://doi.org/10.2514/6.2025-2232}.

\noindent The published version was published by the American Institute of
Aeronautics and Astronautics, Inc., with permission.

\newpage
\setcounter{page}{1}
\maketitle

\begin{abstract}
This study introduces a novel Air Traffic Control (ATC) concept to support self-separation between vehicles in Urban Air Mobility (UAM) corridors. Our proposed scheme involves sharing intended arrival schedules at Constrained Waypoints (CWPs) among UAM operators.
We propose two approaches to assist the arrival scheduling at CWPs by computing the minimum arrival time gap necessary for each pair of vehicles to ensure their safety throughout the flights within the corridor. The first approach considers the minimum separation distance required by the Near Mid-Air-Collision (NMAC) avoidance rules, while the second one is based on the Responsibility-Sensitive Safety (RSS) rules. We demonstrate that the NMAC-rule-based approach can effectively prevent collisions in normal circumstances, where the vehicles adhere to the speed limits of the corridor. However, this approach does not guarantee safety if vehicles exceed the speed limits. Conversely, while the RSS-rule-based approach ensures collision prevention during emergencies when vehicles exceed speed limits, it may require larger arrival time gaps under normal circumstances, which may lead to reduced traffic flow. Our results are confirmed through numerical simulations. 
\end{abstract}

\section{Nomenclature}
\label{sec:Nomenclature}
{\renewcommand\arraystretch{1.0}
\noindent\begin{longtable*}{@{}l @{\quad} l@{}}
UAM &=\quad Urban Air Mobility\\
UAV &=\quad Unmanned Aerial Vehicle\\
ATC &=\quad Air Traffic Control\\
AAM &=\quad Advanced Air Mobility\\
CWP &=\quad Constrained Waypoint\\
ETA &=\quad Estimated Time of Arrival\\
NMAC &=\quad Near Mid-Air-Collision\\
RSS &=\quad Responsibility-Sensitive Safety\\
DAA &=\quad Detect And Avoidance\\
$\corridorLength$ &=\quad length of the considered UAM corridor\\
$\vMin, \vMax$ &=\quad minimum and maximum speed limit of the considered UAM corridor ($\vMin, \vMax > 0$)\\ 
$\aEmergency, \aComfort$ &=\quad maximum and comfortable forward acceleration rates, respectively ($\aEmergency\geq \aComfort > 0$)\\
$\bEmergency, \bComfort$ &=\quad minimum and comfortable braking rates, respectively ($\bEmergency\leq \bComfort < 0$)\\ 
$i$ &=\quad indexing parameter for a vehicle\\ 
$f, r$ &=\quad indexing parameters for a considered pair of vehicles flying in sequence,\\ &\quad~ where $f$ is the front (leading) vehicle and $r$ is the rear (following) one\\ 
$\tiEnter$ &=\quad requested \emph{time instance} by vehicle $i$ to enter the UAM corridor ($\geq 0$)\\
$\durationI$ &=\quad requested \emph{flight time duration} by vehicle $i$ to stay in the UAM corridor  ($> 0$)\\ 
$t$ &=\quad parameter for time instances\\
$\xii(t)$ &=\quad position of the vehicle $i$ at time instance $t$\\ 
$\vi(t)$ &=\quad speed of the vehicle $i$ at time instance $t$\\
$\ai(t)$ &=\quad acceleration rate of the vehicle $i$ at time instance $t$
\end{longtable*}}  


\section{Introduction}

\lettrine{U}{rban} Air Mobility (UAM), which refers to air traffic operational concepts to provide on-demand
or scheduled in a metropolitan
area for both manned and Unmanned Aerial Vehicles (UAVs)~\cite{thipphavong2018urban, vascik2018scaling},
has been receiving attention from industry, academia, and government in the recent years~\cite{thipphavong2018urban}.
As a consequence, the rapidly increasing volume of new UAM operations in metropolitan airspace may overwhelm
the capabilities of current Air Traffic Control (ATC) systems in the near future~\cite{vascik2018scaling}.
The key challenges in high-density UAM operations are 
 congestion reduction and structural
constraints, as most current air transport networks
cannot handle the increasing traffic demand, resulting in congestion and safety issues~\cite{wang2021air}. 
In spite of UAM industry's rapid growth, 
there is a lack of research on UAM
traffic management and flight paths between locations~\cite{bankole2023urban}.

\begin{figure}[bt]
\centering
\includegraphics[width=.7\textwidth]{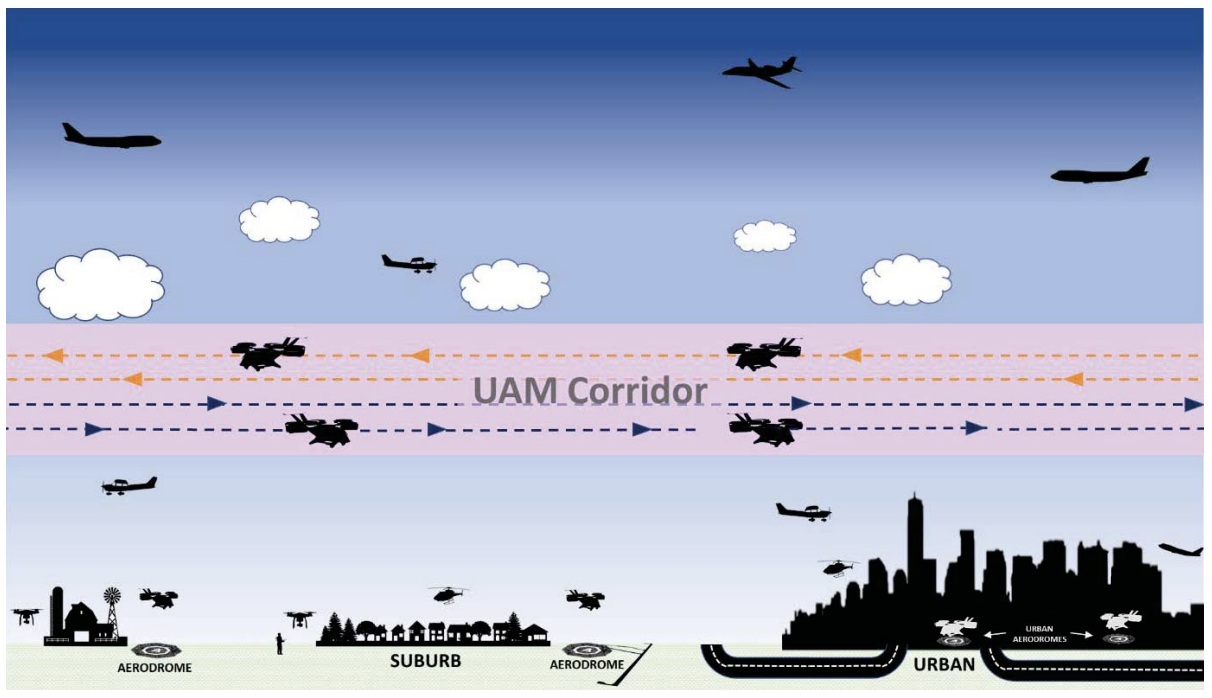}
\caption{UAM Corridor with tracks~(\cite{bradford2020urban}, p.~31).}
\label{fig:conops}
\end{figure}

In FAA's UAM Concept of Operations (ConOps), ~\cite{bradford2020urban, fontaine2023urban}, the Federal Aviation Authority (FAA) proposed the usage of UAM corridors, which are three-dimensional volumes of reserved airspace tracks (see also Fig.~\ref{fig:conops}), as part of the Advanced Air Mobility (AAM) infrastructure.
UAM corridor specifications are designed and controlled under its respective authority, which also has the right to open or close each air corridor and  
may specify the expected performance requirements for vehicles engaging the corridor.
As a result, the usage of UAM corridors provides an air traffic route structure.
The anticipated utilization of UAM corridors involves Unmanned Aerial Vehicles (UAVs) in the near future, including air taxis designed to transport passengers and freight in both rural and metropolitan regions, as well as air ambulances intended to provide rapid first responder services during emergencies~\cite{muna2021air}.
Due to the uncertain movement patterns
of UAVs, UAV traffic is challenging for ATC systems, 
as they must not only address regulatory concerns but also prioritize safety and efficiency~\cite{muna2021air}.

The UAM community and stakeholders are anticipated to collaborate to set operational standards for UAM corridors~\cite{bradford2020urban, bankole2023urban, fontaine2023urban}.
Besides safety concerns, restrictive structures like corridors tend to cause more delays compared to less structured airspace with free flight movement~\cite{bauranov2021designing}.
Consequently, several previous studies have investigated the design of UAM corridors and their ATC systems,  taking into account both safety and traffic flow.
We highlight a few notable examples.
In~\cite{wang2021air},
 Wang et al. presented a macroscopic air traffic assignment model to alleviate the congestion and traffic complexity of UAM transportation network. 
The UAM corridors are represented by a graph, and a traffic assignment problem is formulated using linear dynamical system to minimize congestion and air traffic complexity.
Muna et al. proposed a fundamental model for UAV air corridor design using air cubes, skylanes, intersections, vertiports, and gates~\cite{muna2021air}. 
The authors then introduced a multi-layered air corridor model, which included a detailed analysis of traffic at intersections. They also introduced the concept of air corridor capacity, which is determined by the number of cubes and the gap size between UAVs, and discussed the probability of congestion and collisions within the corridor based on this concept.
In~\cite{jiang2022metrics}, Jiang et al. proposed an  evaluation method for air corridor designs based on traffic data and 
temporal and spatial purpose-specific insights, such as safety 
and environmental impacts. 
In~\cite{bankole2023urban}, Bankole et al. proposed 
two different iterations for a track system in UAM corridors, a horizontal track system and a vertically staggered one, in order to enhance their safety and order.  
In \cite{lee2023airspace}, Lee et al. classified existing airspace structure models into several types and analyzed their strengths and weaknesses. Furthermore, they conducted a quantitative analysis by re-categorizing structured airspace and operational methods and examining their combinations.

Recently, in \cite{wing2022digital}, NASA proposed \emph{Digital Flight}, which is a new operating mode for airspace users that complements and expands upon the existing operating modes of Visual Flight Rules (VFR) and Instrument Flight Rules (IFR). Digital Flight is an operating mode in which flight operations are conducted based on digital information, with operators ensuring flight-path safety through cooperative practices and self-separation, enabled by connected digital technologies and automated information exchange. Digital flight rules (DFR) are expected to meet the emerging needs of this century in the same manner that IFR enabled the expansion of aviation services and airspace use in the last century, capabilities that VFR alone could not support. Some recent studies have proposed ATC concepts in UAM corridors based on DFR. For example, in~\cite{prabhath2023ground}, Prabhath et al. presented
a comprehensive architecture for ground-based communication support for AAM vehicles in UAM corridors.
In~\cite{namuduri2023digital}, Namuduri proposed a digital twin approach for integrated airspace management with several subsystems, including digital flight rules and air corridors.
In~\cite{mccorkendale2024digital}, McCorkendale et al. proposed an approach for collision avoidance strategies using digital traffic lights.


This study introduces a novel concept for Air Traffic Control (ATC) in UAM corridors that aligns well with Digital Flight operating mode and DFR, as we ensure self-separation between vehicles through operational intent sharing among the UAM operators.
Our proposed scheme involves sharing intended arrival schedules at Constrained Waypoints (CWPs) among UAM operators. 
Specifically, we study conditions of Estimated Times of Arrival (ETAs) at the CWPs to ensure safe separation between vehicles through their flight through the corridor.
These ETAs shared the same concept as Required Times of Arrival (RTAs) at airspace boundaries used for self-separation assurance in airborne trajectory management, as discussed in \cite{wing2010comparison}.

Based on this concept,
we propose two approaches to assist the arrival scheduling at CWPs by computing the minimum ETA gap at each CWP for each pair of vehicles to maintain safety throughout their flights in the UAM corridor. 
The first approach focuses on a fixed-length minimum separation distance between vehicles, following the Near Mid-Air-Collision (NMAC) avoidance rules~\cite{johnson2017exploration}.
The second one is based on Responsibility-Sensitive Safety (RSS)~\cite{shalev2017formal}, a well-known rule-based approach that provides a formal safety guarantee to automated driving systems.
Specifically, we consider the RSS rule for safe longitudinal distance in single-lane same-direction traffic scenarios.
Unlike the fixed-length minimum separation distance enforced by the NMAC approach, the RSS rule adjusts the required separation distance based on the vehicles' speed, offering better safety guarantees in some situations.
We present theoretical safety guarantees, followed by numerical simulation results to support our theoretical claims. Furthermore, we present an analysis of the strengths and weaknesses of the two proposed methods.

Departing from most previous research that studies a macroscopic perspective of air traffic, such as the throughput of the corridor or the entire transportation network,
our study focuses on the separation scheme that provides a safety guarantee for each pair of vehicles within a corridor. However, despite this microscopic focus, our method formally ensures the safety of all aerial vehicles operating within the UAM corridor. 

The proposed concept can complement previously existing traffic network models with UAM corridors.
In particular,
these CWPs can also be conceptualized as nodes within a graph representing the traffic network (e.g., the graph in \cite{wang2021air}), while the corridors connecting the CWPs can be considered as edges.   
Our approaches apply to the strategic and tactical planning phases of UAM, which correspond to the planning stages before flight and during flight in the case that an unpredicted event occurs, respectively \cite{causa2022strategic}.  
The approaches work in cooperation with 
Detect And Avoidance (DAA) systems, such as those studied in \cite{johnson2017exploration, shalev2017formal}.
While DAA systems concentrate on evading potential collisions during actual flights using real-time sensor data, we offer an additional layer of safety assurance by focusing on vehicle arrival schedules at CWPs before the flight or their changes during flights.
Moreover, while our primary focus is safety, 
the ETA gaps can also be used to predict the traffic flow.

The remainder of this paper is structured as follows. Section~\ref{sec:UAMcorridor} presents the proposed ATC concept and clarifies the importance of arrival scheduling design at CWPs. Section~\ref{sec:NMAC} introduces an approach to compute Estimated Time of Arrival (ETA) gaps between vehicles based on the NMAC avoidance rules.
Then, in Section~\ref{sec:RSS}, we introduce an alternative ETA gaps computation approach based on the RSS rule for safe distance in single-lane same-direction traffic scenarios.
To demonstrate the effectiveness of these approaches, numerical simulations are presented in Section~\ref{sec:experiment}. 
Finally, we present the conclusion in Section~\ref{sec:conclusion}. 

\section{UAM Corridor Traffic Control using  Constrained Waypoints}
\label{sec:UAMcorridor}

\begin{figure}[bt]
\centering
\includegraphics[width=.7\textwidth]{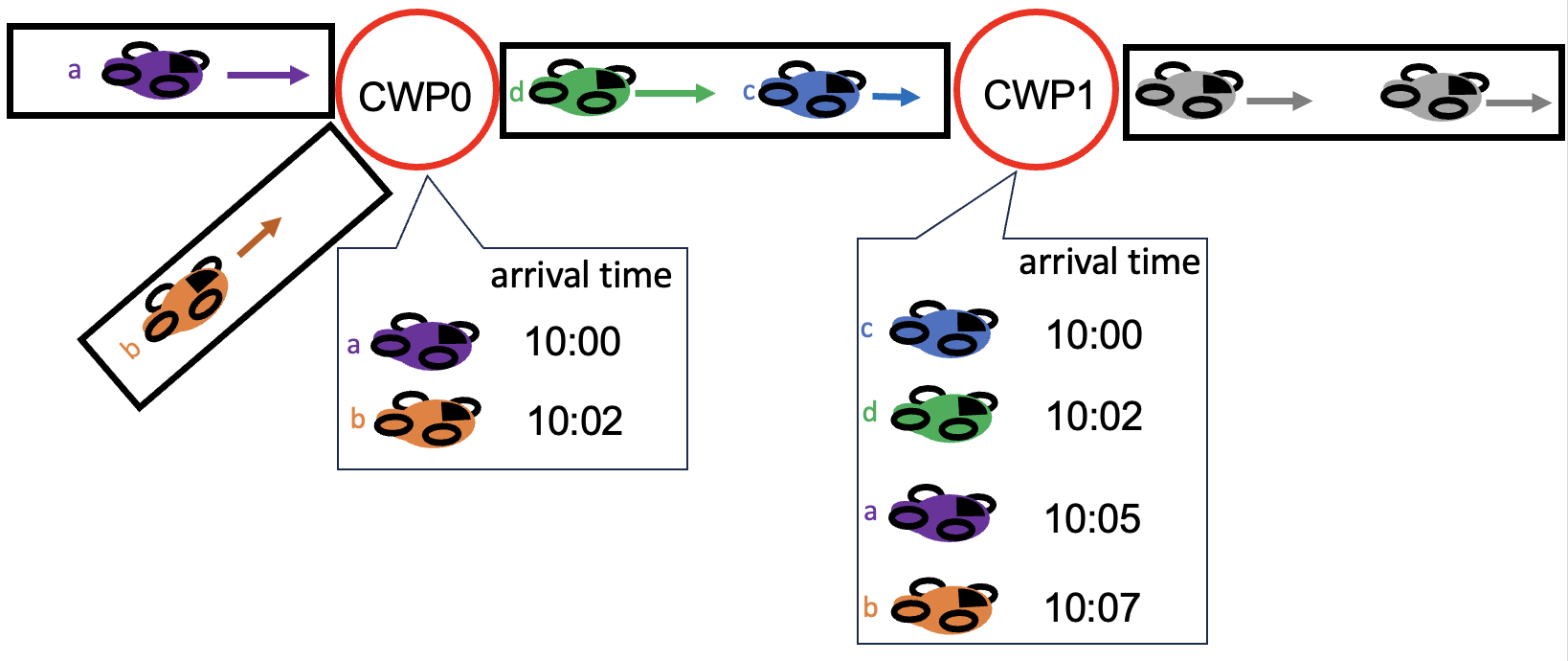}
\caption{A UAM corridor with ETA information.}
\label{fig:corridor}
\end{figure}

In this section, we present an Air Traffic Control (ATC) concept specifically designed for self-separation between vehicles in Urban Air Mobility (UAM) corridors, utilizing the sharing of vehicle intended arrival times at Constrained Waypoints (CWPs). 
As detailed in \cite{bradford2020urban, fontaine2023urban},
UAM corridors designate exclusive airspace for aerial vehicles. An air traffic network typically consists of multiple corridors designated for various phases such as takeoff, cruising, and landing at vertiports. 
As depicted in Fig.~\ref{fig:corridor}, these CWPs segment the air traffic network into multiple UAM corridors. The concept of CWPs can integrate with existing traffic network models featuring UAM corridors. Specifically, these CWPs can be viewed as nodes within a traffic network graph (e.g., the one in \cite{wang2021air}), with the corridors linking the CWPs serving as edges of the graph.

We assume that the UAM corridor is established by the aviation authority and operated and managed by a Provider of Services to UAM (PSU), which supports UAM operators. The PSU may be a single public entity, such as an Air Navigation Service Provider (ANSP), or a group of private operators. As shown in Fig.~\ref{fig:corridor}, we assume that all operators using the corridor submit each vehicle's intended arrival schedule (ETA) at each CWP to the PSU for approval prior to the flight during the strategic planning phase. If the ETA changes during the flight in the tactical planning phase, e.g., due to delays, the PSU will share the updated plan with other operators. In turn, the PSU uses these ETAs to reduce the risk of collisions and airspace congestion.



We focus on a cruising air corridor featuring a single track, such as the corridor between CWP0 and CWP1 as depicted in Figure~\ref{fig:corridor}. To simplify our analysis, we treat the trajectory of UAVs within the corridor as one-dimensional. 
As listed in Section~\ref{sec:Nomenclature}, we denote the position, the speed, and the acceleration rate of a vehicle $i$ at a time instance $t$ using $\xii(t)$, $\vi(t)$, and $\ai(t)$, respectively. The simplified movements of the aerial vehicles adhere to linear motion equations, 
wherein $\vi(t) = d\,\xii(t)/dt$ and $\ai(t) = d\,\vi(t)/dt$.
However, we do not allow vehicles to fly backward by restricting $\vi(t) \geq 0$ at all time instance $t$. 
Two vehicles, denoted as  $f$ and $r$, are considered to have collided if their positions coincide, i.e., $\xf(t) = \xr(t)$, at any given time $t$. 

To ensure safety,
we enforce some restrictions on the UAM corridor.
Let $L$ denote the length of the corridor under consideration. We confine the speed of vehicles operating within the corridor to fall within the specified positive minimum and maximum speed limits, denoted as $\vMin$ and $\vMax$, respectively. Hence, the speed of vehicle $i$ at time $t$ should satisfy the inequality $0 < \vMin \leq \vi(t) \leq \vMax$, as long as vehicle $i$ is flying within the corridor.
Furthermore, in accordance with the Responsibility-Sensitive Safety (RSS) approach~\cite{shalev2017formal}, we assume that the following parameters are defined for the UAM corridor: the maximum forward acceleration rate $\aEmergency$, the comfortable forward acceleration rate $\aComfort$, the minimum braking rate $\bEmergency$, and the comfortable braking rate $\bComfort$. 
In general, we have $0 < \aComfort \leq \aEmergency$ and $\bEmergency \leq \bComfort < 0$.
Principally, $\aComfort$ and $\bComfort$ represent the rates that all considered vehicles can comfortably perform, while $\aEmergency$ and $\bEmergency$ correspond to the extreme rates that the vehicles may potentially execute.
Particularly in a rear-end collision case, 
we usually consider the worst-case scenario where the rear vehicle brakes with the braking rate $\bComfort$ while the front vehicle brakes with the rate $\bEmergency$.
Remark that the parameters  $\aEmergency$, $\aComfort$, $\bEmergency$, and $\bComfort$
can be configured differently for each type of vehicle participating in the UAM corridor. However, for ease of presentation, we assume that these parameters are consistent for all vehicles.

\begin{figure}[bt]
\centering
\includegraphics[width=.7\textwidth]{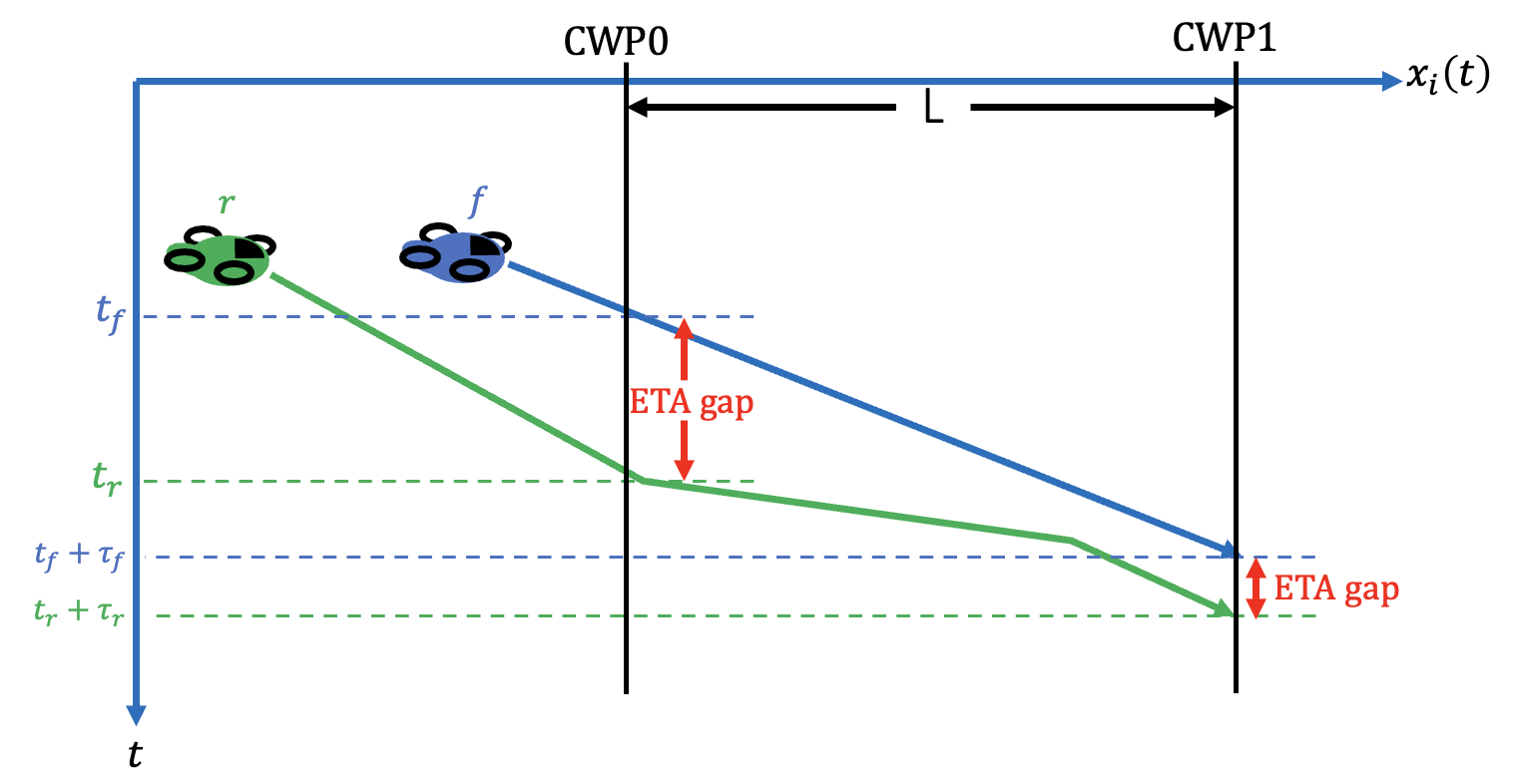}
\caption{An illustrative example for using ETAs to prevent collisions.}
\label{fig:ETA}
\end{figure}

As an illustrative example, let's refer to the spatio-temporal information regarding the trajectories of vehicles $f$ and $r$ in Fig.~\ref{fig:ETA}, where ``$f$'' and ``$r$'' denote ``front'' (leading) and ``rear'' (following) vehicles, respectively. The horizontal axis represents the vehicles' positions ($\xii(t)$, $i \in {f, r}$), while the vertical axis represents the time instances $t$.

The UAM operator of each vehicle $i$ submits the following information to the PSU: 
\begin{itemize} 
\item the schedule for vehicle $i$ to enter the UAM corridor at time $\tiEnter$ at CWP0, and 
\item the requested flight time duration $\durationI$ for the vehicle to spend in the corridor between CWP0 and CWP1 (i.e., vehicle $i$ is scheduled to arrive at CWP1 at time $\tiEnter + \durationI$). 
\end{itemize}
For simplicity, we assume that all vehicles $i$ enter the corridor at position $x_i(\tiEnter) = 0$.
Based on the information submitted to the PSU, we restrict that the planned trajectory of vehicle $i$ satisfies:
\begin{equation} \label{eq:dynamics_restriction}
 x_i(\tiEnter) = 0,~x_i(\tiEnter + \durationI) = \corridorLength, \text{ and }
 x_i(t) \leq x_i(t')  \text{ for all time } t, t' \in [\tiEnter, \tiEnter + \durationI] \text{ such that } t \leq t'.
\end{equation}


Notice that, due to the restricted speed limits imposed on vehicles, the slope of their spatio-temporal trajectories in Fig.~\ref{fig:ETA} is bounded.
Therefore, by restricting suitable arrival times for both vehicles at CWP0 and CWP1, the operators and the PSU can effectively manage the flow of traffic and prevent collisions. Based on this observation, we are interested in the conditions for UAM corridor regulation parameters (e.g., $L, \vMin, \vMax, \trEnter - \tfEnter$) to ensure that all participating vehicles do not collide. 
In this paper, we focus on the following problem. 
\begin{prob}\label{prob:1} 
Consider each pair of vehicles, $f$ and $r$, that are scheduled to enter the corridor in sequence.
Given the speed limit $\vMin$ and $\vMax$, a time instance $\tfEnter$ for vehicle $f$ to enter the corridor, and the flight duration $\durationF$ (\emph{resp.} $\durationR$) that vehicle $f$ (\emph{resp.}  vehicle $r$) requests to spend in the corridor, we need to determine a time instance $\trEnter$ for vehicle $r$ to enter the corridor such that $\xr(t) < \xf(t)$ for all times $t$ within the interval $[\trEnter, \tfEnter + \durationF]$.
\end{prob}

We call $\trEnter - \tfEnter$ the \emph{ETA gap} between $f$ and $r$ for entering the UAM corridor, e.g., at CWP0 in Fig.~\ref{fig:ETA}. 
Notice that if $\tfEnter = 0$, then $\trEnter$ is the ETA gap.
Notice also that solving Problem~\ref{prob:1} also gives us the ETAs for exiting the corridor at CWP1: $\tfEnter + \durationF$ and $\trEnter + \durationR$.

In the following sections, we focus on calculating the minimum ETA gap for each pair of vehicles to ensure their safety throughout their flights within the UAM corridor.
In addition,
to optimize the traffic flow within the corridor, we prefer this ETA gap to be as small as possible.
We present two approaches to compute these minimum arrival time gaps: an NMAC-rule-based approach in Section~\ref{sec:NMAC} and an RSS-rule-based approach in Section~\ref{sec:RSS}.


\section{ETA Gap Computation Based on Near Mid-Air-Collision (NMAC) Avoidance Rules}
\label{sec:NMAC}


According to Near Mid-Air Collision (NMAC) avoidance rules, the established safe separation distance between two manned aircraft is a fixed-length distance, e.g., 152.4m~\cite{johnson2017exploration, muna2021air, jiang2022metrics}. 
Following this regulation, we set the minimum distance for each pair of vehicles to be 
\begin{align}\label{eq:dNMAC}
    \dNMAC(\vMin, \vMax, \mathit{safeD}, \delay) = \mathit{safeD} + (\vMax-\vMin)\, \delay \geq 0,
\end{align}  where $\mathit{safeD}$ is a positive constant for the safe separation distance between the two vehicles, and $\delay$ is a non-negative constant for an upper bound of the delay time. We require this minimum separation distance to be nonnegative.

\begin{figure}[bt]
\centering
\includegraphics[width=.55\textwidth]{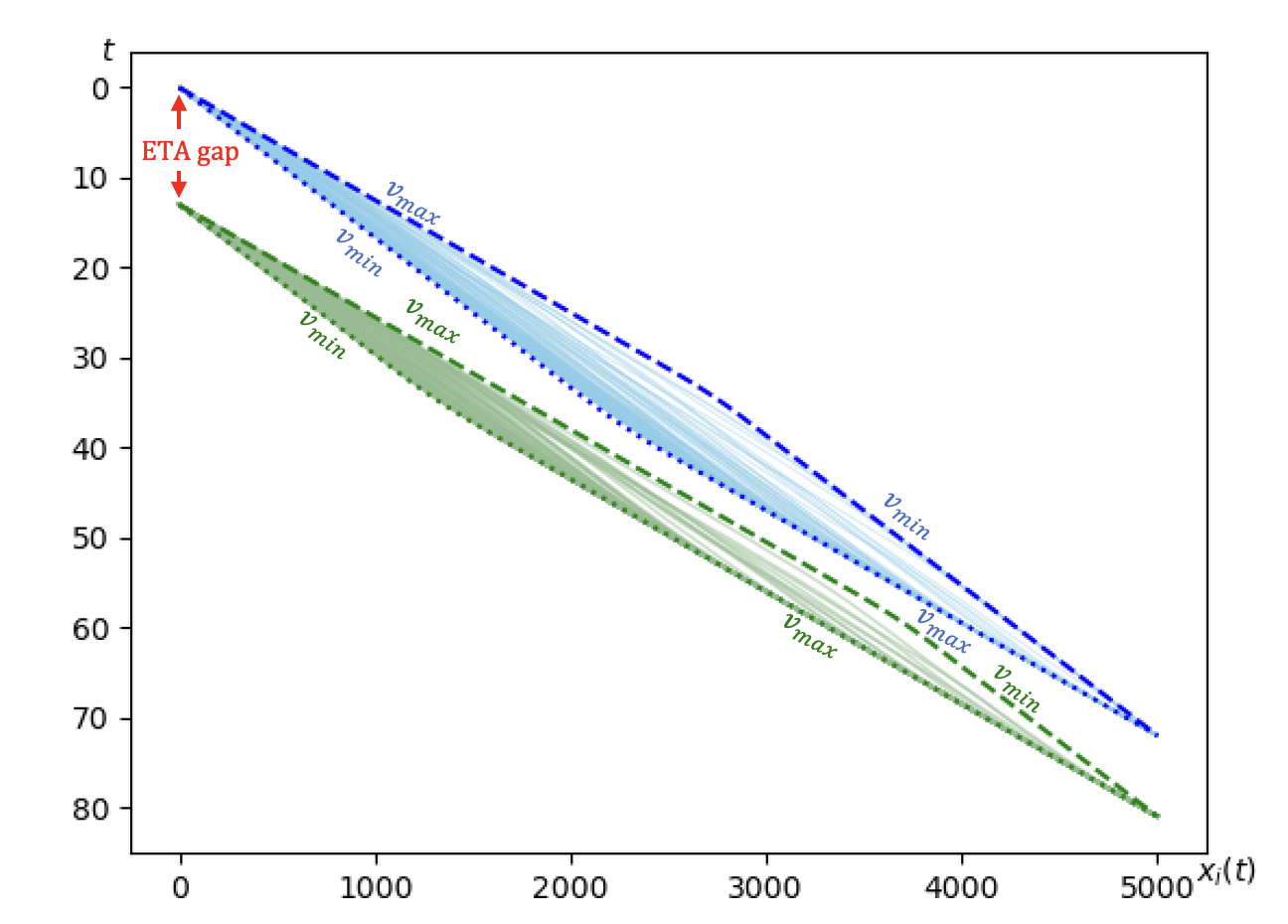}
\caption{A spatio-temporal graph presenting simulated 100 trajectories of vehicles $f$ (blue) and $r$ (green) flying in the corridor in Fig.~\ref{fig:ETA}, where the acceleration rates at each time step are selected uniformly at random from [$\bEmergency$, $\aEmergency$]. 
The dashed (\emph{resp.} dotted) lines represent the border cases computed using the NMAC-rule-based approach with $\mathit{safeD} = 200$,
in which the vehicles enter the corridor with speed $\vMax$ (\emph{resp.} $\vMin$) and later on change its speed to $\vMin$ (\emph{resp.} $\vMax$).
The simulation parameters are provided in Section~\ref{sec:experiment}.
}
\label{fig:NMACnormal}
\end{figure}

Let us consider the scenario depicted in Fig.~\ref{fig:ETA}, where two vehicles, $f$ (front) and $r$ (rear), are scheduled to enter a UAM corridor of length $\corridorLength$. Hence, the PSU must approve the ETAs for both vehicles at constrained waypoints CWP0 and CWP1.
Recall that, due to the restricted speed limits imposed on both vehicles to be within the range $[\vMin, \vMax]$, the slopes of their spatio-temporal trajectories in Fig.~\ref{fig:ETA} are bounded.
In addition, we restrict the planned trajectories to satisfy Eq.~\eqref{eq:dynamics_restriction}. 
Hence, we consider the two boundary cases for trajectories of each vehicle $i \in \{f, r\}$ in the spatio-temporal graph: to enter the corridor with speed $\vMax$ (\emph{resp.} $\vMin$) and later on change its speed to $\vMin$ (\emph{resp.} $\vMax$), as depicted by the dashed (\emph{resp.} dotted) lines in Fig.~\ref{fig:NMACnormal}. It can be easily shown that all other trajectories fall within these boundary cases in the spatio-temporal graph. 
Therefore, ensuring that the two closest border-case trajectories (specifically, when $f$ changes from $\vMin$ to $\vMax$ and $r$ changes from $\vMax$ to $\vMin$) do not collide is sufficient to ensure safety of the pair of vehicle during their flights in the UAM corridor. 

Let's consider a border case in which a vehicle $i$ changes its speed from $v_1$ to $v_2$.
If the vehicle can change its speed instantaneously, we can compute the time $t^*$ for the speed change by the following equations.
\begin{align}
    \corridorLength &= t^* \, v_1 + (\durationI - t^*) \, v_2
    \\
    t^* &=  
    (v_2  \, \durationI - \corridorLength) / (v_2 - v_1)
\end{align}
Hence, we define a function $t_\mathit{i, vChange}$ for the speed-changing time as follows.
\begin{align}
    t_\mathit{i, vChange}(v_1, v_2, \durationI ) =  
    (v_2  \, \durationI  - \corridorLength) / (v_2 - v_1)
\end{align}
However, we also limited the acceleration rate of a vehicle inside the corridor to be in $[\bEmergency, \aEmergency]$. Given an acceleration rate $a \in [\bEmergency, \aEmergency]$, we can compute the time duration that $i$ needs for changing its speed as
\begin{align}
    \tau_\mathit{acc}(v_1, v_2, a) = \frac{v_2 - v_1}{a}.  
\end{align}
Thereby, we have the trajectory function for the border cases as follows.
\begin{align}
   x_i(v_1, v_2, a, \tiEnter, \durationI)(t) = \begin{cases}
        -\infty  & \text{if $t <  \tiEnter$},\\
	v_1 \, (t -  \tiEnter) & \text{if $\tiEnter \leq t < \tiEnter + t_1$},\\
        v_1 \, t_1 + v_1 (t - \tiEnter - t_1) + \frac{1}{2} \, a (t - \tiEnter - t_1)^2  & \text{if $\tiEnter + t_1 \leq t  < \tiEnter + t_2$},\\
        v_1 \, t_1 + \frac{1}{2} (v_1 + v_2) (t_2 - t_1) + v_2 \, (t-\tiEnter-t_2) & \text{if $\tiEnter + t_2 \leq t  < \tiEnter + \durationI$,}\\
        \infty & \text{otherwise.}
		      \end{cases}
\end{align}
where 
\begin{align*}
    t_1 =  
    t_\mathit{vChange}(v_1, v_2, \durationI) - \frac{1}{2}\tau_\mathit{acc}(v_1, v_2, a) 
    \text{ and }
    t_2 =  
    t_\mathit{vChange}(v_1, v_2, \durationI) + \frac{1}{2}\tau_\mathit{acc}(v_1, v_2, a).
\end{align*}

Following the above analysis, we have the following theorem.
\begin{thm}\label{thm:NMAC}
Consider each pair of vehicles, $f$ and $r$, that are scheduled to enter the corridor in sequence. 
    For given $\vMin > 0$, $\vMax \geq \vMin$, $\durationF > 0$,  $\durationR > 0$,  $\tfEnter\geq 0$,
    and $\trEnter\geq 0$ that satisfy
    \begin{align*} 
    \underset{t \in [\trEnter, \tfEnter + \durationF]}{\min} 
	  \xf(\vMin, \vMax, \aEmergency, \durationF)(t) - \xr(\vMax, \vMin, \bEmergency, \durationR)(t) > 0,  
    \end{align*}
    we have
$\xr(t) < \xf(t)$ for all $t 
\in [\trEnter, \tfEnter + \durationF]$.     
\end{thm}

Theorem~\ref{thm:NMAC} establishes a condition of the parameter $\trEnter$ for Problem~\ref{prob:1}, which requires that the trajectories $\xr$ and $\xf$ do not collide at all times inside the corridor. 
To further ensure safety in accordance with NMAC avoidance rules, we enhance this condition by requiring that the distance between these trajectories must exceed the safe distance $\dNMAC(\vMin, \vMax, \mathit{safeD},\delay)$ defined in Eq.~\eqref{eq:dNMAC}. Consequently, we introduce a function for safety margins as follows. 
\begin{align} \label{eq:safetymargin} 
\mathit{safeMargin}_\mathit{NMAC}( \vMin, \vMax, \tfEnter, \trEnter, \durationF, \durationR, \delay) =
\mathit{minD}
- \dNMAC(\vMin, \vMax, \mathit{safeD}, \delay),
\end{align} 
 where, 
\begin{align*}
\mathit{minD} = 
    \underset{t \in [\trEnter, \tfEnter + \durationF]}{\min} 
	  \xf(\vMin, \vMax, \aEmergency, \durationF)(t) - \xr(\vMax, \vMin, \bEmergency, \durationR)(t), 
\end{align*} 
and $\dNMAC(\vMin, \vMax, \delay)$ is given by Eq.~\eqref{eq:dNMAC}. 

By Theorem~\ref{thm:NMAC} and Eq.~\eqref{eq:safetymargin}, we have the following corollary.
\begin{cor}\label{cor:NMAC}
Consider each pair of vehicles, $f$ and $r$, that are scheduled to enter the corridor in sequence.
        For given $\durationF \geq 0$,  $\durationR \geq 0$,  $\tfEnter\geq 0$,
     $\trEnter\geq 0$, and $\delay 
     \geq 0$ that satisfy
    \begin{align*} 
 \mathit{safeMargin}_\mathit{NMAC}( \vMin, \vMax, \tfEnter, \trEnter, \durationF, \durationR, \delay) > 0,  
    \end{align*}
    we have
$\xf(t) - \xr(t) > \dNMAC(\vMin, \vMax, \delay)$ for all $t 
\in [\trEnter, \tfEnter + \durationF]$.   
\end{cor}

Corollary~\ref{cor:NMAC} guarantees that the distance between the two vehicles will always exceed the minimum separation distance mandated by NMAC avoidance rules,
as long as the safety margin specified in Eq.~\eqref{eq:safetymargin} remains positive.

Although the safety-margin constraint is sufficient for ensuring safety, we also aim to minimize the ETA gap $\trEnter$ - $\tfEnter$ to optimize traffic flow. Therefore, we introduce the following optimization problem.
 \begin{align}\label{eq:optimizeNMAC}
\begin{split} 
\underset{\trEnter \in [\tfEnter, \tfEnter + \durationF]}
{\min} \quad &  \trEnter - \tfEnter\\
\textrm{subject to}\quad  & \mathit{safeMargin}_\mathit{NMAC}(\vMin, \vMax, \tfEnter, \trEnter, \durationF, \durationR, \delay) > 0.  
\end{split}  
\end{align}  

By the safety margin constraint and Corollary~\ref{cor:NMAC},
we have that the ETA gap $\trEnter-\tfEnter$ obtained from solving Eq.~\eqref{eq:optimizeNMAC} can prevent collisions, given that both vehicles comply with the speed limits $\vMin$ and $\vMax$.

\begin{figure}[bt]
\centering
\subfigure[]{\includegraphics[width=.42\textwidth]{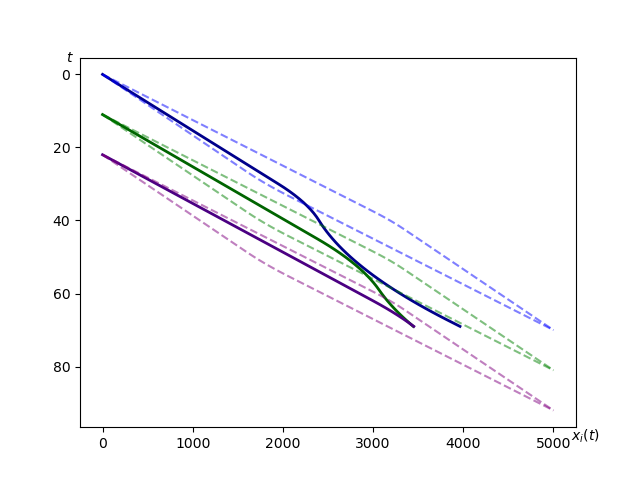}\label{fig:NMACbrake}}
\subfigure[]{\includegraphics[width=.42\textwidth]{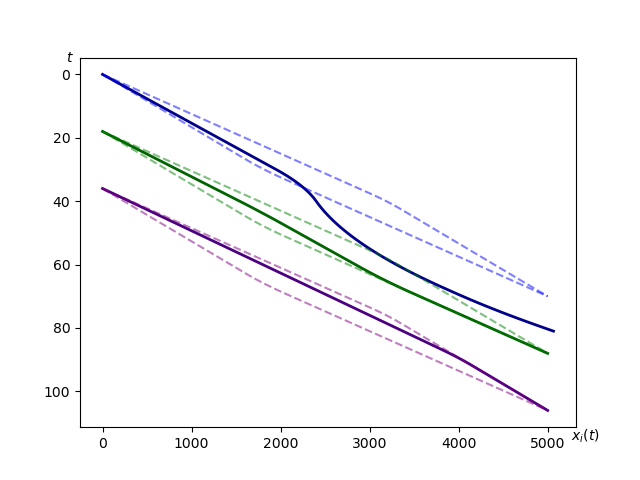}\label{fig:RSSbrake}}
\caption{
Trajectories of three vehicles, \emph{blue}, \emph{green}, and \emph{purple}, entering the UAM corridor in sequence. The leading vehicle, \emph{blue}, performs a sudden brake with braking rate $\bEmergency$ during the time interval $[30\text{s}, 40\text{s}]$. The following two vehicles respond by braking with $\bComfort$ according to (a) the NMAC-rulebased approach with $\mathit{safeD} = 200\text{m}$ and (b) the RSS-rule-based approach. The simulation parameters are provided in Section~\ref{sec:experiment}.
}\label{fig:brake}
\end{figure}

Nonetheless, this approach does not guarantee safety if vehicles surpass the speed limits, as the safety margin constraint directly depends on $\vMin$ and $\vMax$. 
For example, in a simulated instance illustrated in Fig.~\ref{fig:NMACbrake}, we let the leading vehicle, \emph{blue}, suddenly brake with $\bEmergency$ during the time interval $[30\text{s}, 40\text{s}]$, thereby violating the minimum speed limit $\vMin$. 
To ensure safety, we require the following vehicle to be able to avoid the potential collision by applying the comfortable braking rate $\bComfort$.
Following the NMAC avoidance rules, each following vehicle, \emph{green} and \emph{purple}, brakes with $\bComfort$ if its distance from the vehicle in front is not greater than $
 \dNMAC(\vMin=60\text{m/s}, \vMax=80\text{m/s}, \mathit{safeD}= 200\text{m}, \delay=1\text{s})$ (see Section~\ref{sec:experiment} for details). In this instance, \emph{purple} fails to stop in time to avoid colliding with \emph{green}.

In the following section, we introduce an alternative approach that may require a larger ETA gap but can effectively prevent such collisions, as demonstrated in Fig.~\ref{fig:RSSbrake}.

\section{ETA Gap Computation Based on  Responsibility-Sensitive Safety (RSS) Rules}
\label{sec:RSS}

Responsibility-Sensitive Safety (RSS)~\cite{shalev2017formal} is a widely recognized rule-based framework for automated driving systems. RSS is designed to establish safety rules that are precisely defined in mathematical terms. An RSS safety rule consists of an \emph{RSS condition} and its corresponding \emph{proper response}: if the RSS condition is satisfied, executing the corresponding \emph{proper response} ensures safety.

We consider the RSS rule for safe separation distance in single-lane same-direction traffic scenarios with two vehicles: $f$ (front) and $r$ (rear). 
The \emph{minimum safe distance} at a given time instance $t$ is defined as follows:
\begin{align}\label{eq:RSS}
\dRSS(\vf(t) , \vr(t) , \delay)
= \vr(t)  \delay
+ \frac{1}{2} \aEmergency \delay^2
- \frac{(\vr(t) + \aEmergency  \delay)^2}{2 \bComfort}
+\frac{(\vf(t) )^2}{ 2\bEmergency },
\end{align}
where $\delay\geq 0$ represents an upper bound on the delay time. 
To avoid confusion with the minimum safe distance based on the NMAC rules in the previous section, we will refer to the minimum safe distance in Eq.~\eqref{eq:RSS} as the \emph{RSS distance} for the rest of the paper.
 
In this scenario, the \emph{RSS condition} is that the separation distance between vehicles must exceed the RSS distance.
Then, the \emph{proper response} of the following vehicle $r$ is to apply the comfortable braking rate $\bComfort$ within $\delay$ time. The RSS rule ensures that the vehicle $r$ can avoid a rear-end collision with the vehicle $f$ by performing this proper response.
Namely, if the distance between vehicles $f$ and $r$ remains greater than the minimum safe distance, i.e., $(\xf(t) - \xr(t)) - \dRSS(\vf(t), \vr(t), \delay) > 0$, then vehicle $r$ can avoid collision by applying the comfortable braking rate $\bComfort$ in response to vehicle $f$ suddenly braking with the minimum braking rate $\bEmergency$.
Based on this result,
 we have the following corollary, which follows Theorem~2 in~\cite{shalev2017formal}. 
\begin{cor}
    \label{cor:RSS1}
    Consider each pair of vehicles, $f$ and $r$, that are scheduled to enter the corridor in sequence.
    For given $\durationF\geq 0$,  $\durationR\geq 0$,  $\tfEnter\geq 0$,
    and $\trEnter\geq 0$ that satisfy
    \begin{align} \label{eq:RSSminthm1}
\underset{t \in [\trEnter, \tfEnter + \durationF]}{\min} (\xf(t) - \xr(t)) - \dRSS(\vf(t), \vr(t), \delay) \geq 0.
\end{align} 
    we have
$\xr(t) < \xf(t)$ for all $t 
\in [\trEnter, \tfEnter + \durationF]$ if the Detect and Avoidance (DAA) system of the vehicle $r$ can perform the proper response.   
\end{cor}

Here,
Corollary~\ref{cor:RSS1} provides a condition for the parameter $\tfEnter$ for Problem~\ref{prob:1}, ensuring that trajectories $\xr$ and $\xf$ remain collision-free throughout the corridor. For the rest of this section, we assume that the DAA systems of all vehicles can perform the proper response by applying $\bComfort$ when their distance from the vehicle in front is close to violating the RSS condition.


Notice that the RSS distance in Eq.~\eqref{eq:RSS} depends not only on the positions of both vehicles but also on their speeds, as a vehicle traveling at a higher speed requires more time to stop. 
Unlike the NMAC-rule-based approach discussed in the previous section, the RSS rule operates independently of the corridor's speed limits.
It should also be noted that while the RSS rule offers a formal safety assurance for vehicles and can be deployed on DAA systems, its effect on traffic flow is not guaranteed. In this study, we apply the concept of the RSS rule to assist in arrival scheduling at CWPs by computing minimum arrival time gaps to ensure safety, offering an additional layer of safety assurance. Moreover, these arrival time gaps can also be used to estimate traffic flow.

Since the speed change of vehicles inside the corridor is restricted to the allowed acceleration and braking rates,  we will solve the optimization problem in Eq.~\eqref{eq:RSSminthm1} while considering only the feasible speeds of the vehicles.
First, let's examine the situation in which vehicle $i$, $ i \in \{f, r\}$, initiates flight at time $0$ from position $x_0$ with an initial speed $v_0$ and constant acceleration rate $a$, and reaches position $\xii(t)$ at time $t$. Considering the linear motion equations, we have:
\begin{align}
\begin{split}
    \xii(t) &= x_0 + \vi(t) \, t - \frac{1}{2}\, a \, t^2,
    \\
    \frac{\xii(t) - x_0}{t} &= \vi(t) - \frac{1}{2} \, a \, t.
\end{split}
    \label{eq:linear1}
\end{align}

Then, let $v_\mathit{i, avg}^\mathit{pre}$ (\emph{resp.} $v_\mathit{i, avg}^\mathit{post}$) denote the average speed of vehicle $i$ from time $\tiEnter$ to $t$ (\emph{resp.} from time $t$ to $\tiEnter + \durationI$.)
Recall that if we set $a = \aComfort$ (\emph{resp.} $a = \bComfort$), the vehicle accelerates with the highest comfortable acceleration (\emph{resp.} braking) rate. 
By Eq.~\eqref{eq:linear1}, we have the following bounds for $v_\mathit{i, avg}^\mathit{pre}$.
\begin{align}\label{eq:vpre}
    \vi(t) - \frac{1}{2} \, \aComfort \, (t-\tiEnter)) 
    \leq v_\mathit{i, avg}^\mathit{pre} = \frac{x_i(t) - \xii(\tiEnter)}{t-\tiEnter}
    \leq  
    \vi(t) - \frac{1}{2} \, \bComfort \, (t-\tiEnter).
\end{align}
In the same way as the above discussion, we can also derive the bounds for $v_\mathit{i, avg}^\mathit{post}$ and obtain: 
\begin{align}\label{eq:vpost}
    \vi(t) + \frac{1}{2} \, \bComfort \, (\durationI-t + \tiEnter) 
    \leq v_\mathit{i, avg}^\mathit{post} = \frac{\corridorLength - x_i(t) }{\durationI-t + \tiEnter}
    \leq  
    \vi(t) + \frac{1}{2} \, \aComfort \, (\durationI-t + \tiEnter)).
\end{align}

Thus, we can reformulate the optimization problem in Eq.~\eqref{eq:RSSminthm1} by restricting the vehicle speeds 
conditions in Eqs.~\eqref{eq:vpre} and~\eqref{eq:vpost},
and obtain the following function for a safety margin.
\begin{align}
\mathit{safeMargin}_\mathit{RSS}(\tfEnter, \trEnter,& \durationF, \durationR, \delay) = \nonumber
\\
\underset{t \in [\trEnter, \tfEnter + \durationF]}{\text{min}} 
	\quad & (\xf(t) - \xr(t)) - dRSS(\vf(t), \vr(t), \delay) \nonumber
 \\
\textrm{subject to} \quad\quad &\forall  i\in \{f, r\},
\nonumber
\\
        \quad & \xii(t) = v_\mathit{i, avg}^\mathit{pre} \, (t - \tiEnter) = \corridorLength - v_\mathit{i, avg}^\mathit{post} \, (\durationI - t + \tiEnter),
        \nonumber
        \\
        \quad & \vi(t) - \frac{1}{2} \, \aComfort \, (t - \tiEnter) 
    \leq v_\mathit{i, avg}^\mathit{pre}  
    \leq  
    \vi(t) - \frac{1}{2} \, \bComfort \, (t - \tiEnter),  
    \nonumber
        \\
	\quad & \vi(t) + \frac{1}{2} \, \bComfort \,  (\durationI - t + \tiEnter)
    \leq v_\mathit{i, avg}^\mathit{post}  
    \leq  
    \vi(t) + \frac{1}{2} \, \aComfort \, (\durationI - t + \tiEnter),  
    \nonumber
    \\
        \quad & 0 \leq \xii(t) \leq L, 
            \quad \vMin \leq \vi(t) \leq \vMax,
            \nonumber
            \\
        \quad & \vMin \leq  v_\mathit{i, avg}^\mathit{pre} \leq \vMax,
            \quad \vMin \leq  v_\mathit{i, avg}^\mathit{post} \leq \vMax, \label{eq:safetylevel}
\end{align}   
where 
$v_\mathit{i, avg}^\mathit{pre}$ and
$v_\mathit{i, avg}^\mathit{post}$ are as in
Eqs.~\eqref{eq:vpre} and~\eqref{eq:vpost}.

Following the above discussion and Corollary~\ref{cor:RSS1}, we obtain the following theorem.
\begin{thm}
    \label{thm:RSS2}
    Consider each pair of vehicles, $f$ and $r$, that are scheduled to enter the corridor in sequence.
    For given $\durationF\geq 0$,  $\durationR\geq 0$,  $\tfEnter\geq 0$,
    and $\trEnter\geq 0$ that satisfy
    \begin{align*}  
\mathit{safeMargin}_\mathit{RSS}(\tfEnter, \trEnter, \durationF, \durationR, \delay) > 0,
\end{align*} 
    we have
$\xr(t) < \xf(t)$ for all $t 
\in [\trEnter, \tfEnter + \durationF]$.    
\end{thm}

Theorem~\ref{thm:RSS2} ensures that,
as long as the safety margin defined in~Eq.~\eqref{eq:safetylevel} stays positive, the distance between the two vehicles remains greater than the RSS distance.
As a result, the theorem provides a condition for a solution for Problem~\ref{prob:1}. 
Then, our next step is to minimize the ETA gap $\trEnter - \tfEnter$ to optimize traffic flow, achieved by solving the following optimization problem.
 \begin{align}\label{eq:optimizeRSS}
\begin{split} 
\underset{\trEnter \in [\tfEnter, \tfEnter + \durationF]}{\min} 
	\quad &  \trEnter - \tfEnter\\
\textrm{subject to}\quad & \mathit{safeMargin}_\mathit{RSS}(\tfEnter, \trEnter, \durationF, \durationR, \delay) > 0.  
\end{split}  
\end{align}  
 
In the next sections, we perform numerical simulations to demonstrate the effectiveness of the proposed two approaches.

\section{Experimental Results}
\label{sec:experiment}
In this section, we present experimental results to support our theoretical claims in the previous sections.
We perform numerical simulations considering a single-track cruising air corridor with two to eight aerial vehicles entering the corridor sequentially.  
The vehicles are indexed based on their order of entry, with 
$i=0$ for the front-most vehicle.
 
We implement all methods in Python3.12, considering discrete-time dynamics that update every 1s.
Namely, the duration between two simulation time steps is 1s, and the acceleration assigned between the two steps is constant.
For all methods, we set the upper bound for the delay to $\delay = 1$s, corresponding to the duration between two discrete time steps.
We use the following parameters:
the corridor length $\corridorLength = 5000$m, the speed limits $\vMin = 60$m/s and $\vMax = 80$m/s, 
the acceleration limits $\aEmergency=3$m/$\text{s}^2$, $\bEmergency=-4$m/$\text{s}^2$,
$\aComfort=2$m/$\text{s}^2$,  and $\bComfort=-3$m/$\text{s}^2$.

We perform simulations under two scenarios: normal circumstances and airspace-violator.
In normal circumstances, for each vehicle $i$ and each time step $t$,
the acceleration rate $\ai(t)$ is 
uniformly drawn from the range $(\bComfort, \aComfort)$.
However, if necessary, the vehicle $i$ replaces the randomized acceleration $\ai(t)$ with a close one to ensure that $i$ 
adheres to the speed limits $\vMin$ and $\vMax$.
Following the restriction in Eq.~\eqref{eq:dynamics_restriction}, the vehicle also adjusts $\ai(t)$ if it is necessary to ensure it reaches the end of the air corridor at time $\tiEnter + \durationI$, which is its scheduled time to reach the CWP at the end of the corridor (e.g., CWP1 in Fig.~\ref{fig:ETA}), provided that the vehicle speed does not exceed the limits.

For the airspace-violator scenario, we let the vehicles randomize and adjust their acceleration rate $\ai(t)$ in the same way as in normal circumstances, except that for the front-most vehicle $i=0$, we set $a_0(t) = \bEmergency$ for all $t \in [30\text{s},40\text{s}]$. This scenario represents the situation where an airspace violator enters the corridor during the time interval $[30\text{s},40\text{s}]$, forcing the front-most vehicle to ignore the speed limit $\vMin$ and brake to avoid the potential collision with the airspace violator.

For both scenarios,
all vehicles enter the corridor at position $0$m and exit at position $\corridorLength$.
The initial speeds $\vi(0)$ are uniformly selected at random from the range $(\vMin, \vMax)$, while the flight durations $\durationI$ are uniformly selected at random from the range $(65\text{s}, 75\text{s})$.

We perform 1,000 simulations for each of the following methods.
\begin{itemize}
    \item NMAC200, NMAC400, NMAC600, NMAC800, NMAC1000, and NMAC1200 methods use the NMAC-rule-based approach, where the numbers (200, 400, etc.) are the value for the fixed-length distance $\mathit{safeD}$ in Eq.~\eqref{eq:dNMAC}.
    The ETA gaps are computed using the method in Section~\ref{sec:NMAC}.
    The DAA system of each vehicle $i>0$ brakes with $\bComfort$ at time $t$ if $x_{i-1}(t) - \xii(t) \leq \dNMAC(\vMin, \vMax, \mathit{safeD}, \delay)$. 
    \item RSS method uses the RSS-rule-based approach where the ETA gaps are computed using the method in Section~\ref{sec:RSS}.
    The DAA system of each vehicle $i>0$ brakes with $\bComfort$ at time $t$ if $x_{i-1}(t) - \xii(t) \leq
    \dRSS(v_{i-1}(t) , \vi(t) , \delay)$.
\end{itemize}

\begin{figure}[bt]
\centering
\includegraphics[width=.7\textwidth]{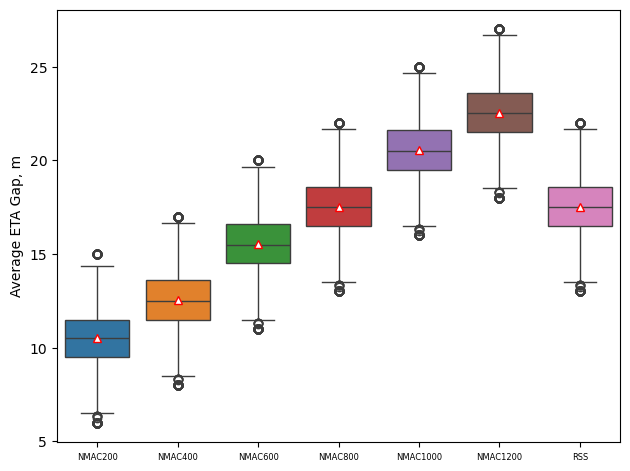}  
\caption{Box-and-whisker plot for the average ETA Gap of each self-separation method.}
\label{fig:boxETA} 
\end{figure}

Fig.~\ref{fig:boxETA} shows the box-and-whisker plot for the average ETA gap of each method, computed by solving the optimization problems in Sections~\ref{sec:NMAC} and \ref{sec:RSS} (Eqs.~\eqref{eq:optimizeNMAC} and \eqref{eq:optimizeRSS}).

\subsection{Performance under Normal Circumstances}
\label{sec:experimentNormal}




\begin{figure}[bt]
\centering
\includegraphics[width=.7\textwidth]{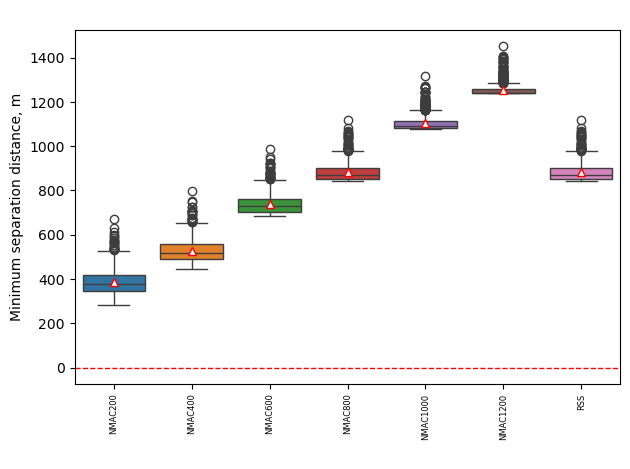}  
\caption{Box-and-whisker plot showing the minimum separation distance between vehicles for each method under normal circumstances, for the case of eight vehicles.}
\label{fig:mindist_normal} 
\end{figure}

First, we conduct simulations for normal circumstances where vehicle speeds always adhere to the limits $[\vMin, \vMax]$. We conduct 1,000 simulations for each of the NMAC200, NMAC400, NMAC600, NMAC800, NMAC1000, NMAC1200, and RSS methods.

Figure~\ref{fig:mindist_normal} shows a box-and-whisker plot showing the minimum separation distance between vehicles for each method under normal circumstances, for the case of eight vehicles.
We do not detect any collision in any simulation for all methods. 
This experimental result confirms that, under normal circumstances in which the 
vehicles adhere to the corridor's speed limits,
all methods can effectively prevent collisions.
This result is expected, as we
compute the ETA gaps by solving the optimization problem in Eq.~\eqref{eq:optimizeNMAC} for the NMAC-rule-based methods (\emph{resp.} Eq.~\eqref{eq:optimizeRSS} for the RSS-rule-based one) subjecting to
the condition where the safe margin in Eq.~\eqref{eq:safetymargin} (\emph{resp.} \eqref{eq:safetylevel}) staying above zero, 
implying that the vehicles must always stay apart.

\begin{table}[bt]
\centering
\caption{Average exit time, s, of the last vehicle from the UAM corridor in normal circumstances} 
\label{tab:tableTimeNormal} 
\includegraphics[width=.8\textwidth]{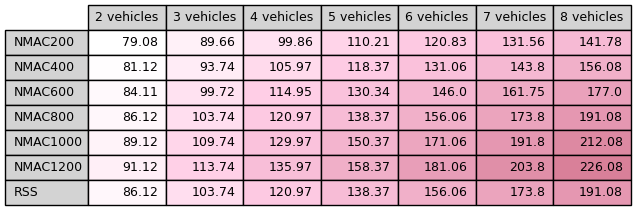}
\end{table}

Table~\ref{tab:tableTimeNormal} presents the average time that the last vehicle exits the UAM corridor. From this table, 
the NMAC-rule-based methods with $\mathit{safeD}\in \{200\text{m}, 400\text{m}, 600\text{m}\}$ seemingly yield better traffic flows 
than the RSS-rule-based method in normal circumstances.

\subsection{Airspace-violator scenario}

\begin{table}[bt]
\centering
\caption{Collision percentage in the airspace-violator scenario} 
\label{tab:tableCollisionViolator} 
\includegraphics[width=.8\textwidth]{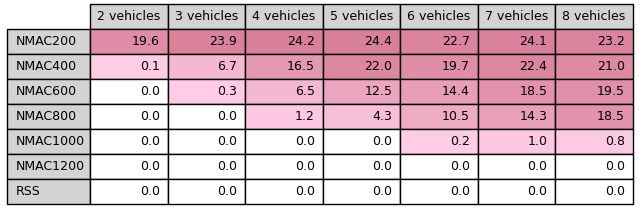}
\end{table} 

\begin{figure}[bt]
\centering
\includegraphics[width=.7\textwidth]{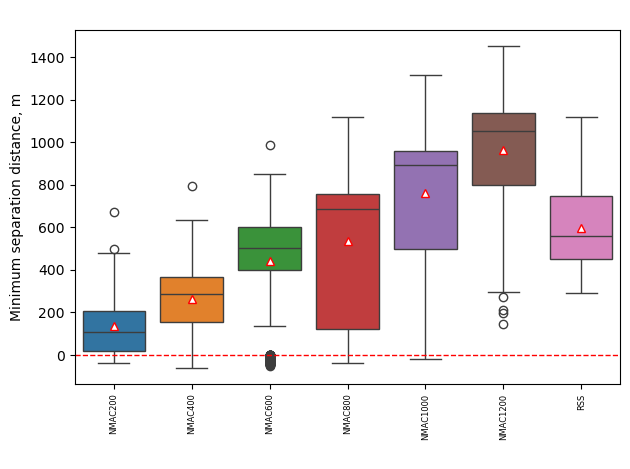}  
\caption{Box-and-whisker plot showing the minimum separation distance between vehicles for each method in the airspace-violator scenario, for the case of eight vehicles.}
\label{fig:mindist_violator} 
\end{figure}

We consider the airspace-violator scenario, which 
 represents the situation where an airspace violator enters the corridor during the time interval $[30\text{s},40\text{s}]$, forcing the front-most vehicle to ignore the speed limit $\vMin$ and brake with the minimum braking rate $\bEmergency$.
 
Table~\ref{tab:tableCollisionViolator} presents the collision percentage in the airspace-violator scenario.
Figure~\ref{fig:mindist_violator} shows a box-and-whisker plot for each method's minimum separation distance between vehicles for the case of eight vehicles.
Unlike in normal circumstances, we detect collisions for all NMAC-rule-based methods except NMAC1200, which uses $\mathit{safeD}=1200$m. However, we do not detect any collision using the RSS-rule-based method.
These results confirm that the RSS-rule-based approach can effectively prevent collisions, even when vehicles violate speed limits.

\begin{table}[bt]
\centering
\caption{
Average exit time, s, of the last vehicle from the UAM corridor in the airspace-violator scenario} 
\label{tab:tableTimeViolator} 
\includegraphics[width=.8\textwidth]{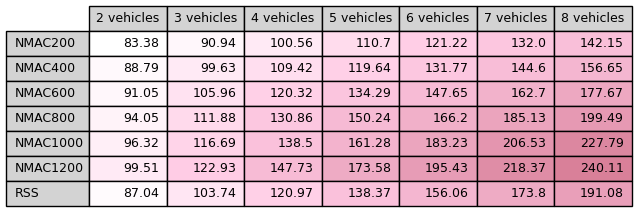}
\end{table}

Table~\ref{tab:tableTimeViolator} presents the average exit time of the last vehicle from the UAM corridor, calculated only from collision-free simulation instances. 
From this table, the RSS-rule-based method seemingly yields better traffic flows 
than the NMAC-rule-based methods with $\mathit{safeD}\geq 800$.

\subsection{Simulation result summary and analysis}
Our simulation results confirm the theoretical guarantees that both the NMAC-rule-based and RSS-rule-based methods effectively prevent collisions under normal circumstances where vehicles respect the corridor's speed limits. However, 
compared to the NMAC-rule-based method with $\mathit{safeD} < 800$m,
the RSS-rule-based method requires larger arrival time gaps (Fig.~\ref{fig:boxETA}) and longer average total flight time (Table~\ref{tab:tableTimeNormal}).
Therefore, the RSS-rule-based method may yield worse traffic flow than the NMAC-rule-based ones with $\mathit{safeD} < 800$m.

In the airspace-violator scenario, our simulation results also confirm the theoretical guarantee that the RSS-rule-based approach can assist the vehicles in avoiding a potential rear-end collision, even when the vehicle speeds exceed the speed limits. On the other hand, the NMAC-rule-based method needs large separation distances and large ETA gaps to achieve the same level of safety, e.g., $\mathit{safeD} \geq 1200$m 
for the scenarios with at least six vehicles entering the corridor. These large separation distances $\mathit{safeD}$ and large ETA gaps may result in worse traffic flow than the RSS-rule-based method. It is a future work to further investigate how our proposed methods affect traffic flows within UAM corridors.

\section{Conclusion}
\label{sec:conclusion}
We presented an Air Traffic Control (ATC) concept designed to support self-separation between vehicles in Urban Air Mobility (UAM) corridors. Our proposed scheme involves sharing intended arrival schedules at Constrained Waypoints (CWPs) among UAM operators.
We proposed two approaches to assist arrival scheduling at (CWPs), by computing the minimum Estimated Time of Arrival (ETA) gap at each CWP for every pair of vehicles, ensuring their safety throughout their flights within the UAM corridor. 
The first approach considers the minimum separation distance between vehicles required by the Near Mid-Air-Collision (NMAC) avoidance rule, while the second approach is based on the Responsibility-Sensitive Safety (RSS) rule for safe separation distance in single-lane same-direction traffic scenarios. 
We provided theoretical safety guarantees for both methods under normal circumstances, which are circumstances where the vehicles respect the corridor's speed limits.
Nonetheless, we only provided a safety guarantee for the RSS-rule-based approach in situations where vehicles may violate the speed limits. 

We performed numerical simulations to support our theoretical claims in normal circumstances and an airspace-violator scenario. 
In normal circumstances where vehicles respect the corridor's speed limits,
our simulation results confirm that both NMAC- and RSS-rule-based approaches effectively prevent collisions. However, the RSS-rule-based approach requires larger arrival time gaps and longer total flight time than the NMAC-rule-based one with small separation distances. Therefore, 
the RSS-rule-based approach may result in worse traffic flow.

The airspace-violator scenario 
 represents the situation where an airspace violator enters the corridor for a time interval, forcing the front-most vehicle to brake while ignoring the speed limit.
In this scenario, our simulation results confirm that the RSS-rule-based approach can assist the following vehicle in avoiding a potential collision. On the other hand, the NMAC-rule-based approach needs a large separation distance and large ETA gaps to achieve the same level of safety, which may result in worse traffic flow than the RSS-rule-based approach.

Future work will involve studying other UAM corridor regulation parameters, conducting a more detailed analysis of the results, and extending the study to encompass other UAM corridor scenarios. We will also investigate how our proposed methods further affect traffic flows within UAM corridors.



\section*{Acknowledgments}
This study is based on results obtained from the ``Realization of Advanced Air Mobility (ReAMo) Project'' by the New Energy and Industrial Technology Development Organization (NEDO).


\bibliography{ref}

\begin{thebibliography}{18}
\newcommand{\enquote}[1]{``#1''}
\providecommand{\natexlab}[1]{#1}
\providecommand{\url}[1]{\texttt{#1}}
\providecommand{\urlprefix}{URL }
\expandafter\ifx\csname urlstyle\endcsname\relax
  \providecommand{\doi}[1]{\discretionary{}{}{}https://doi.org/#1}\else
  \providecommand{\doi}[1]{\discretionary{}{}{}\urlstyle{rm}\url{https://doi.org/#1}}\fi

\bibitem[{Thipphavong et~al.(2018)Thipphavong, Apaza, Barmore, Battiste, Burian, Dao, Feary, Go, Goodrich, Homola et~al.}]{thipphavong2018urban}
Thipphavong, D.~P., Apaza, R., Barmore, B., Battiste, V., Burian, B., Dao, Q., Feary, M., Go, S., Goodrich, K.~H., Homola, J., et~al., \enquote{Urban air mobility airspace integration concepts and considerations,} \emph{2018 Aviation Technology, Integration, and Operations Conference}, 2018, p. 3676.

\bibitem[{Vascik and Hansman(2018)}]{vascik2018scaling}
Vascik, P.~D., and Hansman, R.~J., \enquote{Scaling constraints for urban air mobility operations: Air traffic control, ground infrastructure, and noise,} \emph{2018 aviation technology, integration, and operations conference}, 2018, p. 3849.

\bibitem[{Wang et~al.(2021)Wang, Delahaye, Farges, and Alam}]{wang2021air}
Wang, Z., Delahaye, D., Farges, J.-L., and Alam, S., \enquote{Air traffic assignment for intensive urban air mobility operations,} \emph{Journal of Aerospace Information Systems}, Vol.~18, No.~11, 2021, pp. 860--875.

\bibitem[{Bankole(2023)}]{bankole2023urban}
Bankole, I., \enquote{Urban Air Mobility (UAM) Flight Path: Literature Review and Conceptual Design of UAM Corridor Virtual Lane System using “Tracks”,} 2023.

\bibitem[{Bradford(2020)}]{bradford2020urban}
Bradford, S., \enquote{Urban Air Mobility Concept of Operations v1.0,} Federal Aviation Administration, Office of NextGen, 26 June 2020.
\newblock \urlprefix\url{https://nari.arc.nasa.gov/sites/default/files/attachments/UAM_ConOps_v1.0.pdf}.

\bibitem[{Fontaine(2023)}]{fontaine2023urban}
Fontaine, P., \enquote{Urban Air Mobility Concept of Operations v2.0,} Federal Aviation Administration, Office of NextGen, 26 April 2023.
\newblock \urlprefix\url{https://www.faa.gov/sites/faa.gov/files/Urban%20Air%20Mobility%20%28UAM%29%20Concept%20of%20Operations%202.0_1.pdf}.

\bibitem[{Muna et~al.(2021)Muna, Mukherjee, Namuduri, Compere, Akbas, Moln{\'a}r, and Subramanian}]{muna2021air}
Muna, S.~I., Mukherjee, S., Namuduri, K., Compere, M., Akbas, M.~I., Moln{\'a}r, P., and Subramanian, R., \enquote{Air corridors: Concept, design, simulation, and rules of engagement,} \emph{Sensors}, Vol.~21, No.~22, 2021, p. 7536.

\bibitem[{Bauranov and Rakas(2021)}]{bauranov2021designing}
Bauranov, A., and Rakas, J., \enquote{Designing airspace for urban air mobility: A review of concepts and approaches,} \emph{Progress in Aerospace Sciences}, Vol. 125, 2021, p. 100726.

\bibitem[{Jiang et~al.(2022)Jiang, Peng, Bulusu, Poliziani, Chatterji, and Sengupta}]{jiang2022metrics}
Jiang, X., Peng, X., Bulusu, V., Poliziani, C., Chatterji, G., and Sengupta, R., \enquote{A Metrics-based Method for Evaluating Corridors for Urban Air Mobility Operations,} \emph{2022 IEEE International Smart Cities Conference (ISC2)}, IEEE, 2022, pp. 1--7.

\bibitem[{Lee et~al.(2023)Lee, Ahn, Choi, Chin, and Jang}]{lee2023airspace}
Lee, U.-J., Ahn, S.-J., Choi, D.-Y., Chin, S.-M., and Jang, D.-S., \enquote{Airspace Designs and Operations for UAS Traffic Management at Low Altitude,} \emph{Aerospace}, Vol.~10, No.~9, 2023, p. 737.

\bibitem[{Wing et~al.(2022)Wing, Lacher, Ryan, Cotton, Stilwell, Maris, and Vajda}]{wing2022digital}
Wing, D., Lacher, A., Ryan, W., Cotton, W., Stilwell, R., Maris, J., and Vajda, P., \enquote{Digital flight: A new cooperative operating mode to complement VFR and IFR,} NASA/TM–20220013225, 1 Sep. 2022.
\newblock \urlprefix\url{https://ntrs.nasa.gov/api/citations/20220013225/downloads/NASA-TM-20220013225.pdf}.

\bibitem[{Prabhath et~al.(2023)Prabhath, Sun, Jayaweera, Manu, Kakaraparty, Pallav, Mahbub, Mandapaka, Rochi, Meraz et~al.}]{prabhath2023ground}
Prabhath, K., Sun, X., Jayaweera, S.~K., Manu, D., Kakaraparty, K., Pallav, S., Mahbub, I., Mandapaka, J.~S., Rochi, S.~D., Meraz, M., et~al., \enquote{Ground-Based Communication Support for Air Corridors,} \emph{2023 IEEE 34th Annual International Symposium on Personal, Indoor and Mobile Radio Communications (PIMRC)}, IEEE, 2023, pp. 1--6.

\bibitem[{Namuduri(2023)}]{namuduri2023digital}
Namuduri, K., \enquote{Digital Twin Approach for Integrated Airspace Management with Applications to Advanced Air Mobility Services,} \emph{IEEE Open Journal of Vehicular Technology}, 2023.

\bibitem[{McCorkendale et~al.(2024)McCorkendale, McCorkendale, Kidane, and Namuduri}]{mccorkendale2024digital}
McCorkendale, Z., McCorkendale, L., Kidane, M.~F., and Namuduri, K., \enquote{Digital Traffic Lights: UAS Collision Avoidance Strategy for Advanced Air Mobility Services,} \emph{Drones}, Vol.~8, No.~10, 2024, p. 590.

\bibitem[{Wing et~al.(2010)Wing, Prevot, Murdoch, Cabrall, Homola, Martin, Mercer, Hoadley, Wilson, Hubbs et~al.}]{wing2010comparison}
Wing, D.~J., Prevot, T., Murdoch, J.~L., Cabrall, C.~D., Homola, J.~R., Martin, L.~H., Mercer, J.~S., Hoadley, S.~T., Wilson, S.~R., Hubbs, C.~E., et~al., \enquote{Comparison of Airborne and Ground-Based Function Allocation Concepts for NextGen Using Human-In-The-Loop Simulations,} \emph{10th AIAA Aviation Technology, Integration and Operations (ATIO) Conference}, 2010.

\bibitem[{Johnson et~al.(2017)Johnson, Petzen, and Tokotch}]{johnson2017exploration}
Johnson, S.~C., Petzen, A., and Tokotch, D., \enquote{Exploration of detect-and-avoid and well-clear requirements for small UAS maneuvering in an urban environment,} \emph{17th AIAA Aviation Technology, Integration, and Operations Conference}, 2017, p. 3074.

\bibitem[{Shalev-Shwartz et~al.(2017)Shalev-Shwartz, Shammah, and Shashua}]{shalev2017formal}
Shalev-Shwartz, S., Shammah, S., and Shashua, A., \enquote{On a formal model of safe and scalable self-driving cars,} \emph{CoRR abs/1708.06374}, 2017.

\bibitem[{Causa et~al.(2022)Causa, Franzone, and Fasano}]{causa2022strategic}
Causa, F., Franzone, A., and Fasano, G., \enquote{Strategic and tactical path planning for urban air mobility: Overview and application to real-world use cases,} \emph{Drones}, Vol.~7, No.~1, 2022, p.~11.

\end{thebibliography}

\end{document}